\documentclass[aps,prb,twocolumn,amsmath,amssymb]{revtex4-1}
\usepackage{amsmath,amssymb,amsfonts,bm}
\usepackage{graphicx}
\usepackage{color}
\usepackage{bbold}
\usepackage{graphicx}
\usepackage{dcolumn}
\usepackage{epstopdf}
\usepackage{color}
\usepackage{epsfig}
\usepackage{titletoc}
\usepackage{url}
\usepackage{hyperref}
\usepackage{verbatim}

\newcommand{\bit}{\begin{itemize}}
\newcommand{\eit}{\end{itemize}}

\newcommand{\ba}{\begin{align}}
\newcommand{\ea}{\end{align}}
\newcommand{\be}{\begin{equation}}
\newcommand{\ee}{\end{equation}}
\newcommand{\bi}{\begin{itemize}}
\newcommand{\ei}{\end{itemize}}
\newcommand{\lf}{\left(}
\newcommand{\ri}{\right)}

\newcommand{\id}{\mathbb{1}}

\newcommand{\n}{\nonumber \\ }

\begin{document}
\title{Long-range Kitaev Chains via Planar Josephson Junctions}
\author{Dillon T. Liu, Javad Shabani, and Aditi Mitra}
\affiliation{Center for Quantum Phenomena, Department of Physics, New York University, New York, NY, 10003, USA}
\date{\today}

\begin{abstract}
We show how a recently proposed
solid state Majorana platform comprising
a planar Josephson junction proximitized to a 2D electron gas (2DEG) with Rashba spin-orbit coupling and Zeeman field
can be viewed as an effectively
one-dimensional (1D) Kitaev chain with long-range pairing and hopping terms. We highlight how the couplings of the 1D system may be tuned
by changing experimentally realistic parameters.
We also show that the mapping is
robust to disorder by computing the Clifford pseudospectrum index in real space for the long-range Kitaev chain across several topological phases.
This mapping opens up the possibility of using current experimental setups to explore 1D topological superconductors with non-standard, and tunable
couplings.
\end{abstract}

\maketitle

\section{Introduction}\label{intro}
Topological superconductivity is currently being studied in great detail and in a wide variety of settings\cite{AliceaRev, FlensbergMajorana, BeenakkerMajorana, StanescuMajorana, AguadoMajorana, LutchynMajorana}. The experimental platforms involve combinations of topological materials,
proximity-effect induced superconductivity, and strong spin-orbit coupled semiconductors and
nanowires\cite{SauProp, AliceaProp, LutchynWire, OppenProp, Rokhinson2012, BergProp, YazdaniProp, BrauneckerProp, Loss2013, BenaProp, Loss2016, MTJProp}.
The interest in these systems is due to the existence of Majorana bound states. These are zero-energy modes localized at the boundary of
topological superconductors, and might be used as
a topologically robust means of processing quantum information \cite{KitaevTQC, NayakRMP, Bravyi1, Bravyi2, Freedman}.

It has been recently proposed that an experimental platform based on a planar Josephson junction comprising a 2DEG with spin-orbit coupling,
can host Majorana modes\cite{BergHalperin, Flensberg}. This set-up is well characterized
experimentally \cite{Yacoby, Rokhinson, Marcus, ShabaniPRB2016, Henri, Marcus3, Fabrizio,Kaushini}, and may have the advantage of
better identifying the origin of zero-bias
conductance peaks \cite{Stanescu2018}. More importantly, this experimental set-up
may be used to fabricate more elaborate networks. Such scalability is essential for carrying out complex operations with
multiple Majoranas, including braiding, an essential component for topological quantum computation.

On the other hand, the Kitaev chain is arguably the simplest theoretical setting in which one can study topological superconductivity and
Majorana modes \cite{KitaevChain}. The Kitaev chain consists of spinless fermions in one dimension with $p$-wave superconductivity. Due to the interests
in seeking Majorana modes, much experimental and theoretical work has been conducted to realize the Kitaev model and related generalizations.
Moreover the long-range Kitaev chain shows a significantly richer phase diagram with intriguing distinctions from the
nearest-neighbor Kitaev chain \cite{NiuLRK, SenLRK,LeporiLRK, LeporiLRK2, MassiveEdgeModes, AlecceLRK, DuttaLRK, CaiLRK}. Due to this, there are many
proposals for
realizing long-range Kitaev chains in physical systems, including schemes to induce long-range coupling via Floquet periodic driving of
static nearest-neighbor Kitaev chains, ultracold atomic and trapped ion systems with photonic coupling,
and solid state systems \cite{FloquetLRK1, FloquetLRK2, Monroe, Zoller, Biercuk, PientkaLRK, LeporiSS}.

In this paper we show that the new experimental platform employing planar Josephson junctions  described above\cite{BergHalperin,Flensberg} is
equivalent to realizing a long-range Kitaev chain. We discuss this mapping, present numerical evidence for the long-range nature of the system,
and examine the robustness to disorder by studying topological phase transitions via a novel real space invariant, the Clifford pseudospectrum
method \cite{HastingsLoringEPL, HastingsLoring, Loring, Fulga}.

The paper is organized as follows.
In Sec.~\ref{background}, we review the Majorana platform proposal and discuss its topological classification according to the
Altland-Zirnbauer (AZ) scheme\cite{AZ, KitaevClass, LudwigClass}. In Sec.~\ref{mapping}, we present details of the mapping between
the planar 2D and 1D systems. Lastly, in Sec.~\ref{realspace}, we discuss the real space topological invariant for a disordered system,
and present the numerical results. An appendix provides some additional discussion on the real space topological invariant constructed out of the Clifford pseudospectrum.
We end with an outlook in Section~\ref{Summary}.

\section{Topological Superconductivity in Planar Josephson Junctions}\label{background}

We review the proposal for obtaining Majorana bound states in a 2D solid state setting.~\cite{BergHalperin,Flensberg}
Consider a 2DEG with strong spin-orbit coupling (e.g. InAs), an in-plane magnetic field, and $s$-wave superconducting leads that proximitize the 2DEG
to leave a quasi one-dimensional channel. The Hamiltonian is,
\begin{align}
{\cal H} &= \frac{1}{2}\Psi^\dag\cdot H_\text{BdG} \cdot \Psi,\\
H_\text{BdG} &= \lf -\frac{\nabla^2}{2m} - \mu\ri\tau_z - i\alpha a \lf\partial_x \sigma_y - \partial_y \sigma_x \ri \tau_z\n
&\quad\quad+ E_Z\lf y\ri \sigma_x + \Delta \lf y\ri\tau_+ + \Delta^\star\lf y\ri\tau_-,
\end{align}
where $\Psi = \lf c_{\uparrow}, c_{\downarrow},c^\dag_{\downarrow},-c^\dag_{\uparrow}\ri^{\intercal}$. $H_\text{BdG}$ includes
Rashba spin-orbit coupling of strength $\alpha$, an in-plane Zeeman field $E_Z$ along the longitudinal direction, and two proximitized superconducting
leads. $\sigma_i,\tau_i$ are the Pauli matrices acting in spin and particle-hole space respectively, and $a$ is a lattice spacing.
\begin{figure}
	\includegraphics[width=0.45\textwidth]{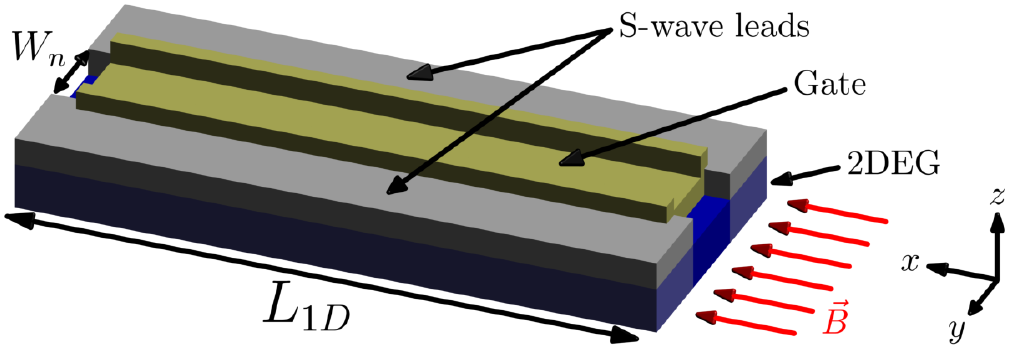}
	\caption{Schematic of 2D Majorana platform proposed in \cite{BergHalperin, Flensberg}. A 2DEG with strong spin-orbit coupling is contacted with
$s$-wave superconducting leads that have a phase difference $\phi$ and an in-plane longitudinal magnetic field, $\vec{B}$. The top gate can be used to tune the chemical potential. Majorana edge modes appear at
either end of the 1D normal channel.}
	\label{fig:illustration1}
\end{figure}

We note that the Zeeman field is different in the leads because of the proximity effect and different $g$-factors. Additionally, we allow a phase difference, $\phi$, between the two superconducting leads. Notably, in this system, there will be topological regimes for all values of $\phi$ away from $\phi = 0$, depending on other parameters in the system (e.g. $E_Z$), see Fig.~1c of Ref.~\cite{BergHalperin}. As indicated in Fig.~\ref{fig:illustration1}, the proximitizing leads give rise to a quasi-1D normal channel of length $L_{1D}$ and width $W_n$. Following \cite{BergHalperin}, we take $\Delta \lf y\ri = \Delta e^{i\text{sgn}\lf y\ri\phi/2}\theta\lf \lvert y \rvert - W_n/2\ri$ and $E_Z\lf y\ri = E_{Z,s}\theta\lf\lvert y\rvert-W_n/2 \ri + E_{Z,n}\theta\lf W_n/2 - \lvert y\rvert\ri$. The presence of many experimental knobs is useful for controlling the topological phase of the system, especially if one wishes to manipulate a network of Majorana modes. However, we will focus on just one of these, the chemical potential, $\mu$.

Current experimental progress makes it feasible to study the physics discussed
here. In 2D InAs, the difficulty of creating strong uniform coupling to a superconductor was recently resolved by growing the superconducting material
(e.g. Al) in situ by molecular beam epitaxy \cite{ShabaniPRB2016}. A hard superconducting gap, measured by tunneling through a quantum point contact,
indicated an intimate coupling between materials \cite{Marcus}.  Gated-Josephson junctions similar to Fig.~\ref{fig:illustration1} are routinely fabricated
with $I_cR_n$ product of $\sim \Delta_{\rm Al}$ \cite{Kaushini}, and signature of Majorana fermions have been observed on similar heterostructures \cite{Fabrizio, Henri}.

Past proposals have focused on realizing Majorana modes, manipulating the modes, and maximizing the topological gap which protects them, but our interest is on
another aspect, namely to highlight this set-up as a tunable Kitaev chain with long-range couplings.

{\sl Topological Classification:}
We now discuss the topological classification of this system via the AZ scheme \cite{AZ, KitaevClass, LudwigClass}.
Imposing periodic boundary conditions along the longitudinal $\hat{x}$ direction, and open boundary conditions in the transverse $\hat{y}$ direction, and
going into momentum space in the $\hat{x}$ direction,
\begin{align}
{\cal H} &= \sum_{k_x} \Psi_{k_x}^\dag H_\text{BdG} \lf k_x, y \ri \Psi^{\phantom{dag}}_{k_x}.
\end{align}
Denoting $\mathcal{K}$ as complex conjugation, and identifying two anti-unitary transformations,~\cite{BergHalperin, SatoEffectiveTSym} particle-hole by
$U_C \mathcal{K}$, with $U_C \equiv \tau_y \sigma_y$, and time reversal by $U_T\mathcal{K}$, with
$U_T \equiv -\lf y \to -y\ri$,
\begin{align}
U_C H_\text{BdG}\lf k \ri U_C^{-1} &= -H^\star_\text{BdG}\lf -k \ri,\\
U_T H_\text{BdG}\lf k \ri U_T^{-1} &= H^\star_\text{BdG}\lf -k \ri,
\end{align}
where $k$ replaces the notation $k_x$.
The two anti-unitary symmetries also imply the chiral symmetry
$U_S = U_C U_T$,
\begin{align}
U_S H_\text{BdG}\lf k \ri U_S^{-1} &= -H_\text{BdG}\lf k \ri.
\end{align}
Since $\left(U_{C,T}\mathcal{K}\right)^2=1$,
$H_\text{BdG} \lf k \ri$ is in the AZ symmetry class BDI , and therefore has a topological invariant in $\mathbb{Z}$\cite{KitaevBDI}.

{\sl Topological Invariant:}
In systems with a chiral symmetry, there is an established method for computing the $\mathbb{Z}$ invariant \cite{TewariSauInvariant}. This method applies generally to multiband systems in AZ class BDI, and has previously been used to study quasi-one dimensional Rashba spin-orbit coupled nanowires \cite{TewariSauInvariant, TSSDSInvariant}, in which the systems are multibanded because of the finite width of the confining quantum well. In contrast, in the systems considered in our work, multiple bands arise from the bound states in the junction. Furthermore, the mirror symmetry in our system allows the chiral symmetry to remain exact in the presence of interband coupling.

First, we rotate $H_\text{BdG}$ to the
basis in which $U_S$ has the form $\Gamma = \tau_z \otimes \id\otimes\id$, where the two unit matrices appearing in the tensor product live in $\sigma$ and position
space. In this basis, $H_\text{BdG}$ is block off-diagonal,
\begin{align}
\Gamma &= \begin{bmatrix}
    \id & 0\\
    0 & -\id\\
\end{bmatrix}, H_\text{BdG} = \begin{bmatrix}
    0 & h\lf k \ri\\
    h^\dag\lf k \ri & 0\\
\end{bmatrix}.
\end{align}
\begin{figure}
	\includegraphics[width=0.45\textwidth]{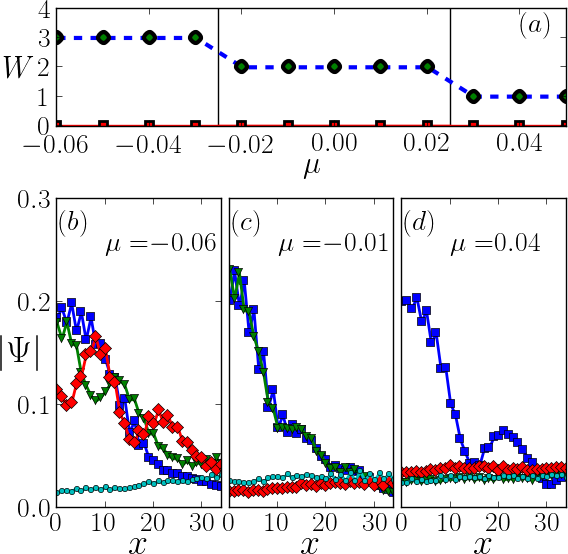}
	\caption{$(a)$ Comparison of the topological invariant computed via winding number in 2D system (dashed line) and via the real space invariant in the
1D model (circles) defined in Eq.~\eqref{Index}. The fluctuations across disorder realizations are also plotted (boxes), but are very close to zero. $(b-d)$ The amplitude of four positive energy modes closest to zero for a finite 1D system in the three topological phases shown in the top panel. The amplitudes are averaged over disorder realizations ($N_R = 100$) and show that there are as many edge modes as the invariant predicts.}
	\label{fig:invariant_mems}
\end{figure}

A Hamiltonian with a gap at zero energy has non-zero determinant and therefore the following quantities are well-defined,
\begin{align}
\alpha\lf k \ri &= \frac{\det h\lf k \ri }{\lvert \det h\lf k \ri \rvert} \equiv e^{ i\theta\lf k\ri },\\
W &= \frac{1}{2\pi i}\int \frac{d\alpha\lf k \ri }{\alpha\lf k \ri}.
\end{align}

Because of the periodic boundary conditions, $W$ is a quantized winding number and may only change if the gap closes. Thus $W$ is the topological invariant in $\mathbb{Z}$. Further, by the bulk-boundary correspondence, $W$ gives the number of Majorana modes at the boundary of the bulk for which the invariant is computed. In this case, the bulk is the quasi-1D, normal channel and the zero-energy modes will be forced to the ends of the channel.

In the following, the numerical parameters used for the 2D system have been chosen to be experimentally realistic for InAs devices \cite{Flensberg, Kaushini}. We take, in units of the 2D hopping $t = 1/(2m^\star a^2)$, $E_{Z,n} = 0.02$, $E_{Z,s} = 0.001$, $\alpha = 0.2$, $\Delta = 0.02$, and $\phi = \pi/2$, with $a = 10$nm, $m^\star = 0.03m_e$. We consider a normal channel of width $W_n = 8$ sites on which $\Delta = 0$ and superconducting leads with $250$ sites in the transverse direction. When putting the system on a finite wire, we take $L_{1D} = 800$ sites. Disorder strength is $D/2 = 1$.

Fig.~\ref{fig:invariant_mems} shows the invariant for a tight-binding model of the bulk 2D system as one tunes through
several topological phase transitions. Fig.~\ref{fig:winding_angle} shows the winding angle explicitly
for a single choice of parameters. One can see that for the chosen parameters, the angle
wraps around the unit circle three times, leading to $W=3$.

\section{Mapping to 1D Kitaev chain with long-range couplings} \label{mapping}
We now discuss the mapping from the 2D system to the 1D Kitaev chain. Typically, the Kitaev chain is discussed with nearest-neighbor coupling terms and has a topological invariant in $\mathbb{Z}_2$ for class D. By including further neighbor terms, the invariant can take values fully in $\mathbb{Z}$ \cite{YatesMitra, YatesLemonikMitra}.
Thus the simplest example of a Hamiltonian in topological class BDI can be written as
\begin{align}
H\lf k \ri &= \vec{d}\cdot \vec{\tilde{\tau}} = d_z\lf k \ri \tilde{\tau}_z + d_y \lf k\ri \tilde{\tau}_y,
\end{align}
where $\tilde{\tau}$ are Pauli matrices (not to be confused with $\tau,\sigma$ that are Pauli matrices in particle/hole and spin space, respectively) and the symmetries of class BDI require that there is no component of $\vec{d}$ along $\tilde{\tau}_x$. Consequently, the winding of the unit vector, $\vec{d}/|\vec{d}|$, is a topological invariant and takes values in $\mathbb{Z}$.
\begin{figure*}
	\includegraphics[width=0.8\textwidth]{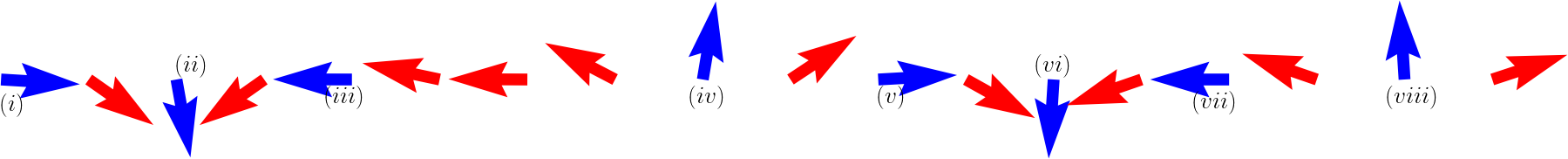}
	\includegraphics[width=\textwidth]{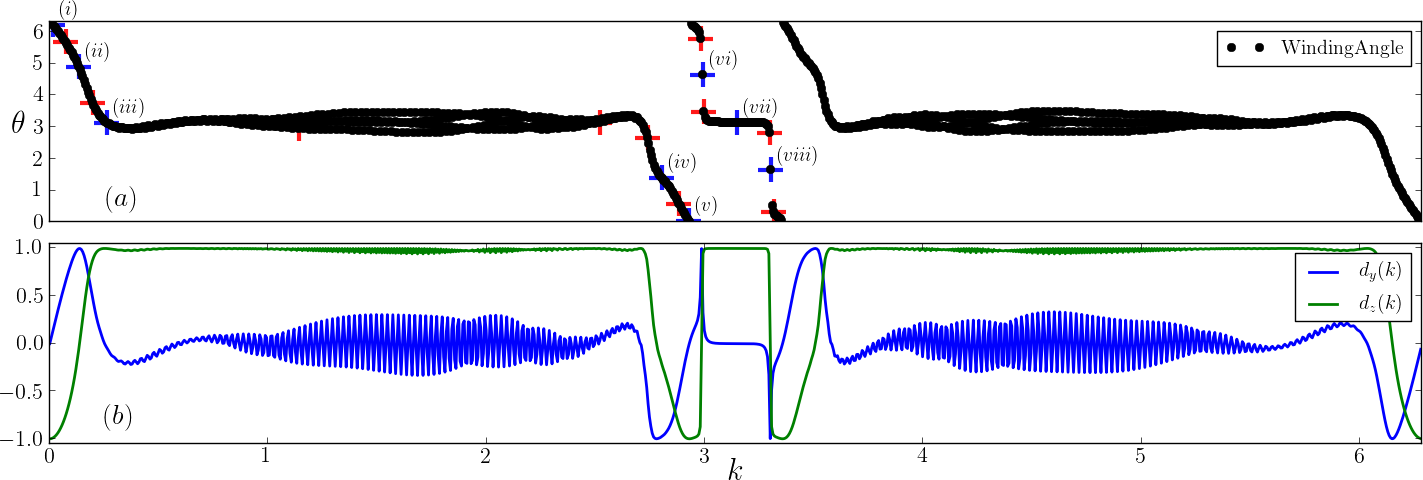}
	\caption{(a) Winding angle $\theta \lf k \ri$ extracted from a 2D system. This angle winds around the unit circle three times to give an invariant $W=3$. Several points are extracted and labeled with roman numerals $(i)-(viii)$ illustrating the winding angle. (b) Components $d_z\lf k \ri, d_y\lf k \ri$, defining a 1D Hamiltonian, as been derived from a 2D system. These differ dramatically from simple sinusoidal functions, which implies that the 1D system is not a nearest-neighbor model, but instead includes long-range hopping and pairing terms.}
	\label{fig:winding_angle}
\end{figure*}
In particular, for the 1D system, the winding may be explicitly computed by
\begin{align}
W &= \frac{1}{ 2\pi }\int d\theta \lf k \ri,
\end{align}
where $\theta\lf k \ri$ is the angle defined by the components of $\vec{d}$. If one breaks the effective time reversal symmetry (TRS), a $\tilde{\tau}_x$ term may appear in $H$. Such TRS breaking may arise intentionally (as in Ref. \cite{Flensberg}) by differing superconducting gaps in the leads or unintentionally through disorder which does not obey $y\to-y$. However, broken TRS will reduce the invariant from $\mathbb{Z}$ to $\mathbb{Z}_2$, still leading to a non-trivial topological phase.

We can now make a straightforward connection with the invariant described above for the 2D system. To map the full 2D system to the 1D system, we use $\alpha\lf k \ri$
above from the 2D system to define components $d_z\lf k \ri, d_y\lf k \ri$. These components then characterize a 1D Hamiltonian. Specifically, we write
\begin{align}
d_z\lf k\ri &\equiv \cos\left[\theta\lf k \ri \right],\\
d_y\lf k\ri &\equiv \sin\left[\theta\lf k \ri \right].
\end{align}

In a Kitaev chain with only nearest-neighbor hopping and pairing, these terms are simple and well-known, $d_z\lf k \ri = \cos\lf k\ri, d_y\lf k\ri = \sin \lf k\ri$.
Systems that include further neighbor terms will have higher-harmonic contributions to $d_z, d_y$. We show that when the mapping above is carried out,
the Hamiltonian which characterizes the effective 1D system includes long-range hopping and pairing terms. The initial indication that this is the case
comes from functional forms of $d_z\lf k\ri, d_y\lf k \ri$, which deviate significantly from simple sinusoids. We illustrate this in Fig.~\ref{fig:winding_angle},
where $d_z,d_y$ clearly include higher-harmonic contributions.
We will examine these contributions more concretely and their implications below.

The long-range couplings of the 1D problem may be intuitively thought of as coming from integrating out closely spaced modes
residing in the transverse direction of the 2DEG.

\section{Long-range Kitaev chain: Real Space and Disorder} \label{realspace}
We now consider the long-range Kitaev chain in real space to study the extent of the long-range coupling explicitly. After mapping to an effective 1D system,
we place the long-range Kitaev chain on a finite wire, and examine the spatial dependence of the coupling terms.
We then introduce disorder to the 1D model and find that the topological properties are unchanged. To study the topology in real space for a system without translation invariance, we use methods discussed in \cite{HastingsLoringEPL, HastingsLoring, Loring, Fulga} which rely on the Clifford pseudospectrum.

\begin{figure}
	\includegraphics[width=0.45\textwidth]{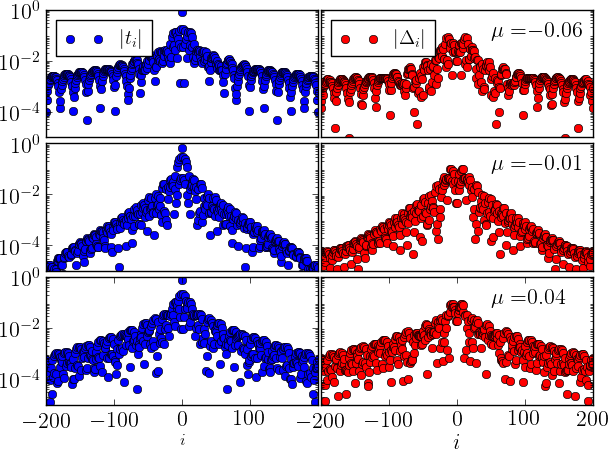}
	\caption{(left panels) Hopping and (right panels) pairing parameters ($t_i, \Delta_i$ resp.) are plotted as functions of the separation, $i$, between sites involved. These parameters are shown for three different values of 2D chemical potential, which correspond to three distinct topological regimes. The presence of significant terms beyond nearest-neighbors is evident and there is a noticeable difference in the decay at large separations for the various topological regimes.}
	\label{fig:dz_dy_single_params}
\end{figure}

In order to obtain the spatial behavior, we compute the discrete inverse Fourier transform of $d_z\lf k \ri,d_y\lf k \ri$ and obtain the hopping and pairing terms as a function of $x$. If $d_z,d_y$ are computed at $L_{1D}$ discrete values of $k$, the finite wire correspondingly has $L_{1D}$ sites. The resulting coefficients $t_i, \Delta_i$ give the $i$th neighbor (up to $L_{1D}/2$)  hopping and pairing coupling strengths. In Fig.~\ref{fig:dz_dy_single_params}, we show an example of the coupling coefficients as a function of the separation between the sites that they couple. Fig.~\ref{fig:dz_dy_single_params} displays the long-range nature of these couplings.

In real space, the Hamiltonian for the 1D model involving site $j$ is thus
\begin{align}
H_j &= \mu_{1D}n_j + \sum_{\substack{0 < \lvert i\rvert \leq L_{1D}/2,\\ 0 < j+i< L_{1D}}} t_{i} c_{j}^\dag c_{j+i} + \Delta_{i}c_{j}^\dag c_{j+i}^\dag + \text{h.c.},
\end{align}
where $n_j \equiv c_j^\dag c_j$, the full Hamiltonian is $H = \sum_{j=0}^{L_{1D}-1}H_j$, $\mu_{1D} \equiv t_0/2$ (where $t_0$ is the coefficient of the on-site terms, $c_j^\dag c_j$) is the effective chemical potential of the 1D system, and open boundary conditions have been imposed.
\begin{figure}
	\includegraphics[width=0.45\textwidth]{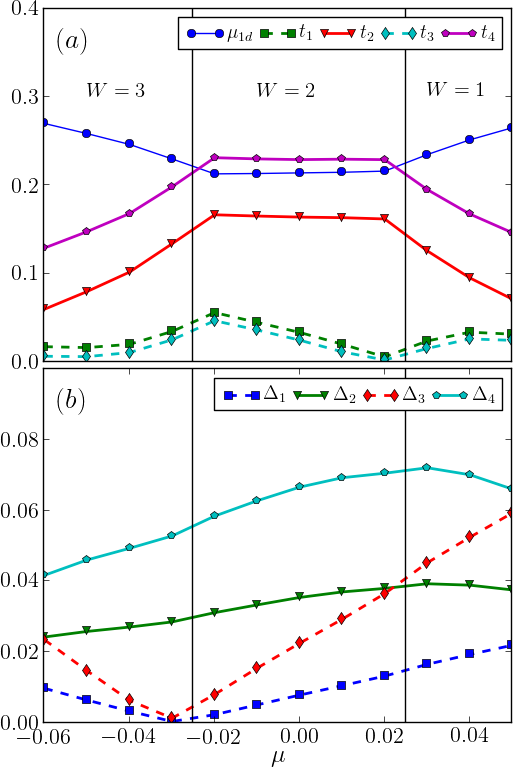}
	\caption{$(a)$ Hopping and $(b)$ pairing parameters for the 1D model vary as the underlying 2D system is tuned through several topological phase transitions, as indicated in Fig.~\ref{fig:invariant_mems}. In these panels, we have included the first four nearest neighbor coupling terms.}
	\label{fig:1D_params}
\end{figure}

{\sl Real Space Topological Signature:}
The most prominent methods and tools for identifying topological phases and their invariants have relied on momentum space computations, as in the case of the winding number discussed above\cite{TewariSauInvariant}. However, these methods cannot be used when systems lack translation invariance, as in the case of aperiodic or disordered systems. For such systems without translation invariance, Clifford pseudospectrum methods have proven exceptionally useful for identifying topological features \cite{HastingsLoringEPL, HastingsLoring, Loring, Fulga}.

The real space invariants depend on the AZ classification scheme. For class BDI, the index which characterizes the topology of a system is given by the signature of a combination of the matrices $H_\text{BdG}, \Gamma$, and $X$, where $X$ is the position operator normalized by the system size.
\begin{align}
\text{Index} &\equiv\frac{1}{2}\text{sig}\left[X\Gamma + H_\text{BdG}\right] \label{Index},
\end{align}
where $\text{sig}\left[M\right]$ is the difference between the number of positive and negative eigenvalues of a matrix $M$ counting degeneracies. The index changes when an eigenvalue of $X\Gamma + H_\text{BdG}$ crosses zero. When the index is non-zero, there is a simultaneous eigenvector of $X$ and $H_\text{BdG}$ which is localized in space and near zero energy \cite{Fulga}. Consequently, a non-zero index indicates a topologically nontrivial phase in the bulk and determines the number of zero-energy modes. The Appendix gives a more detailed discussion of this invariant.

{\sl Numerical Results:}
We now turn to the results from studying finite, disordered wires in 1D. We demonstrate that the topological invariant is preserved through, first, the mapping to 1D and, second, the breaking of translation invariance by the presence of disorder. This is shown in Fig.~\ref{fig:invariant_mems}, where the topological invariant has been determined in two ways. First, we compute the usual topological invariant determined by the winding number in the bulk, translation invariant 2D system. Second, we compute the real space signature defined above for the disordered, finite wire and find that this agrees with the former across several topological phase transitions.

In Fig.~\ref{fig:invariant_mems}, the fluctuations (across disorder realizations) of the real space signature are also shown, but are typically zero or very small. This indicates the robustness of the topological properties of the system to disorder. Only in the presence of very strong disorder can the topological superconducting phase be destroyed and taken to a trivial phase. On the other hand, in the presence of very weak disorder, we find that the system size must be increased for the real space signature method to match the bulk invariant. This is due to the longer localization length (in the case of weaker disorder) which undermines the Clifford pseudospectrum method that requires sufficiently spatially localized eigenstates\cite{Loring}.

We introduce disorder in the 1D system chemical potential. Concretely, we add the disorder via $\mu_{1D} + \delta$ where $\delta \in \left[-D/2, D/2\right]$ for disorder strength $D$, and $\mu_{1D}$ is the effective chemical potential for the wire. While this is arguably the simplest kind of disorder to include, from the perspective of the 2D system, it may not be the most physical type of disorder, as we argue below.

In the 2D system, a single physical parameter can be tuned to bring the system through a topological phase transition, as we have seen here with $\mu$. However, in the language of 1D system, a change in the physical chemical potential does not necessarily correspond to an analogous change in $\mu_{1D}$. Instead, we see that the topological phase transitions occur because of collective changes in the 1D hopping and pairing parameters. We plot some of these parameters in Fig.~\ref{fig:1D_params} as the 2D parent system is tuned through topological phase transitions to give a sense of the behavior of the 1D system. Not only does the 1D chemical potential vary, but all of the 1D parameters show smooth variation within a given topological regime. However, these parameters have kinks when the topological invariant changes. Consequently, to include the full effects of disorder on the 2D system, it may be necessary to include disorder in all of the other 1D system parameters (i.e. in the hopping and pairing terms), in addition to $\mu_{1D}$. However given the robustness of the topological phases to disorder, the results for the mapping between 2D and 1D systems are expected to still hold.

Another means of studying the collective evolution of the 1D parameters under changes to a physical parameter of the 2D system is examining the full set of coupling terms as a function of separation. Fig.~\ref{fig:dz_dy_single_params} shows the hopping and pairing terms as a function of the distance between sites they couple and shows a power-law decay at large separations. We have attempted to fit the decay, but it is substantially more subtle than a simple power-law. Prior studies on long-range Kitaev chains have indicated that changes in the strength of the power-law decay of coupling can affect the topological properties of a system \cite{LeporiLRK, LeporiLRK2, MassiveEdgeModes, AlecceLRK, DuttaLRK, CaiLRK, LeporiSS}.

In class BDI, the bulk-boundary correspondence and the $\mathbb{Z}$ topological invariant indicate that we can have more than one Majorana edge mode. This prediction is borne out explicitly in the 1D system as seen in Fig.~\ref{fig:invariant_mems}, where we have shown the four modes closest to zero energy. As the topological invariant changes from $W = 3\to2\to1$ in Fig.~\ref{fig:invariant_mems}, we expect the corresponding number of zero-energy modes pinned to the edge of the wire. In Fig.~\ref{fig:invariant_mems}, we show that when the topological invariant is $3$, there are three zero-energy modes pinned to the edge, one of which disappears as the topological invariant decreases. The non-zero modes are flat after disorder averaging because Anderson localization dictates that the states will be localized and uniformly distributed throughout the system.

\section{Summary and Outlook} \label{Summary}
We have elucidated the relationship between a recent proposal to realize Majorana bound states in a 2D system and a generalized Kitaev chain, the latter
with long-range hopping and pairing.
By employing a  novel characterization of topological invariants in real space based on the Clifford pseudospectra, we have shown the robustness to this 2D to 1D
mapping to disorder. Interesting avenues of research are opened up given that a current day solid state experiment can be mapped to an exotic model in 1D.
These are to explore the role of
driving, strong interactions, and disorder, features that are difficult to explore in a bulk 2D system. Moreover this work also highlights the importance of interpreting
transport in the solid state device in the language of Majorana
modes of a long-range model and the possibility of realizing some of the new features of long-range Kitaev chains \cite{LeporiLRK, LeporiLRK2, MassiveEdgeModes, AlecceLRK, DuttaLRK, CaiLRK,LeporiSS}. One expects that finite temperature and gate voltage dependence of the conductance will crucially depend on the
long-range features of the model.

{\sl Acknowledgments:}
We thank Yonah Lemonik and Daniel Yates for helpful discussions. DTL and AM were supported by the US Department of Energy, Office of Science,
Basic Energy Sciences, under Award No.~DE-SC0010821. JS was supported by the US Army Office of Research and US Air Force Office of Scientific Research Young Investigator Award.

\newpage
\begin{appendix}
\setcounter{equation}{0}
\begin{widetext}
\section*{Appendix}\label{Appendix}
In this appendix, we provide a basic introduction to the Clifford pseudospectrum invariant used in the main text. We suggest Refs.~\onlinecite{Loring, Fulga} to readers seeking a formal derivation. Here we use the one-dimensional Kitaev chain to illustrate the invariant. Our Hamiltonian has $L$ sites, ${\cal H} = \sum_{xx'} \Psi^\dag_x H \lf x, x' \ri \Psi_{x'}$, where $\Psi_x = \lf c_x, c^\dag_x\ri^{\intercal}$ and
\begin{align}
H &= -\mu\tilde{\tau}_z\otimes \id - 2t\tilde{\tau}_z\otimes \left[\textbf{hop}\right] - i\Delta\tilde{\tau}_y\otimes \left[\textbf{pair}\right] - \tilde{\tau}_z\otimes \left[\textbf{W}\right],
\end{align}
where $\tilde{\tau}_i$ are the Pauli matrices and $\left[\textbf{hop}\right]$ and $\left[\textbf{pair}\right]$ are $L\times L$ matrices corresponding to nearest neighbor hopping and pairing as shown below:
\begin{align}\nonumber
\left[\textbf{hop}\right] =\begin{bmatrix}
    0       &    1         &    0         &    0         &\dots\\
    1       &    0         &    1         &    0         &\dots\\
    0       &    1         &    0         &    1         &\dots\\
    0       &    0         &    1         &    0         &\dots\\
    \vdots  &    \vdots    &    \vdots    &    \vdots    &\ddots
\end{bmatrix},
\left[\textbf{pair}\right] =\begin{bmatrix}
    0       &    1         &    0         &    0         &\dots\\
    -1      &    0         &    1         &    0         &\dots\\
    0       &   -1         &    0         &    1         &\dots\\
    0       &    0         &   -1         &    0         &\dots\\
    \vdots  &    \vdots    &    \vdots    &    \vdots    &\ddots
\end{bmatrix}.
\end{align}
$\left[\textbf{W}\right]$ is a diagonal matrix with uniformly distributed entries $\delta \mu \in \left[ -w,w\right]$ corresponding to disorder in the chemical potential. We also consider the coordinate operator, $X = \id \otimes \tilde{x}$, where $\tilde{x}$ is a $L\times L$ matrix with diagonal entries corresponding to position, and normalized so that the entries $x$ range over $-1/2 \leq x \leq 1/2$.
It is also convenient to define $\Gamma = \text{Diag}\lf\id,-\id\ri$. This is the chiral operator where $H\Gamma=-\Gamma H$ and
$X \Gamma=\Gamma X$.

The first step in identifying the real space invariant is to look for operators that almost commute. In the presence of disorder, and in one dimension, $X,H$ are such a set of operators. A convenient way~\cite{Loring} to construct the simultaneous (approximate) eigenvectors of $X,H$ with eigenvalues $\lambda_x,\lambda_H$ is via the Clifford pseudospectrum, $\Lambda\lf X,H\ri$ defined as 
\begin{align}
\Lambda \lf X, H\ri &= \left\{ \bm{\lambda} \in \mathbb{R}^2 | B_{\bm{\lambda}} \lf X, H \ri \,\text{is singular} \right\}, \\
\bm{\lambda} &= \lf \lambda_x, \lambda_H\ri,\\
B_{\bm{\lambda}}\lf X, H\ri &= \Gamma_a \otimes \lf X-\lambda_x\ri + \Gamma_b \otimes \lf H-\lambda_H\ri.
\end{align}
Above, by singular one means that the matrix cannot be inverted due to zero eigenvalues. 
$\Gamma_a, \Gamma_b$ are a Hermitian representation of ${\cal C}\ell_2 \lf\mathbb{C}\ri$ such that $\Gamma_a\Gamma_b = -\Gamma_b\Gamma_a, \Gamma_i^2 = 1, \Gamma_i^\dag = \Gamma_i$. Clearly, if $H$ has a spectral gap, and the states are localized, then ${\bf 0} \not\in \Lambda\lf X,H\ri$.
The only way for the topological classification to change is when ${\bf 0} \in \Lambda\lf X,H\ri$.

Let us choose $\Gamma_a=\sigma_x,\Gamma_b =\sigma_y$. Then,
\begin{equation}
B_{\bm 0}(X,H) = \begin{pmatrix} 0 && X- i H \\ X+i H && 0\end{pmatrix}
\end{equation}
Our goal is to study $B$, where a singular structure implies a topological phase transition. Since the two non-zero elements are 
conjugates of each other, 
one may identify the singular structure by simply studying one 
of the off-diagonal elements $X+iH$. It is convenient to cast this in a Hermitian form by defining $N'=(X+iH)\Gamma$. Since $\Gamma$ is
unitary, it does not modify the norm of $X+iH$, and hence its singular stucture.
A further unitary
rotation~\cite{Loring} can transform $N'$ to
\begin{eqnarray}
N= X\Gamma + H
\end{eqnarray}

Since a topological phase transition implies a zero eigenvalue of $N$, we can simply study the signature of $N$. A change in the signature
implies a level crossing through zero. Moreover, if $H$ is even dimensional, the invariant becomes,
\begin{eqnarray}
W= \frac{1}{2}\text{sig}\left[X\Gamma + H\right]
\end{eqnarray}

We now show plots of the energy spectrum, pseudospectrum (eigenvalues of $N$), and eigenstates of $N = X\Gamma + H$ for the Kitaev chain defined above. First, we consider the energy spectrum and pseudospectrum as a function of $\mu$ for a finite system with open boundary conditions and parameter values $L = 120,t = 1, \Delta = 1, w = 0.5$.
\begin{figure}
	\includegraphics[width=0.48\textwidth]{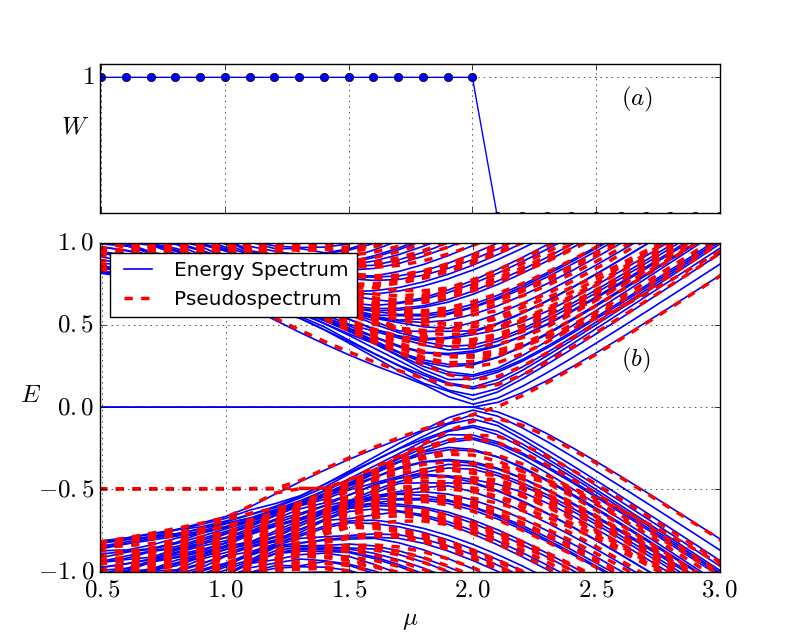}
	\includegraphics[height = 0.36\textwidth]{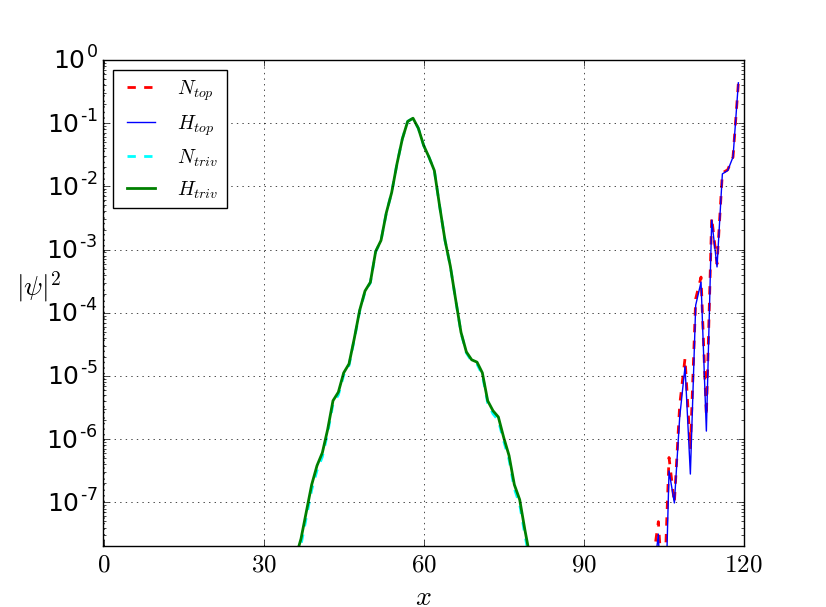}
	\caption{(a) Topological invariant, $W$ defined as half the signature of $N=X\Gamma + H$, plotted for varying $\mu$. (b) (solid blue) Energy spectrum and (dashed red) pseudospectrum of Kitaev chain for varying $\mu$. One can clearly see that the pseudospectrum crosses zero around $\mu = 2$. This is the usual topological phase transition that corresponds to the appearance of edge modes at zero energy. (right) Probability density of eigenstates with eigenvalue closest to zero of $N, H$ for two values of $\mu$ corresponding to being deep in the trivial ($\mu = 3.0$) and topological ($\mu = 1.0$) regimes. The probability density computed from $N,H$ are almost indistinguishable, with the topological edge-mode clearly visible.}
	\label{fig:pseudospec}
\end{figure}

In Fig.~\ref{fig:pseudospec}, we can see that the pseudospectrum includes zero (or something very close to zero) when $\mu$ is around $2$. This agrees with the phase transition in the Kitaev chain between the trivial and topological regimes. We can also see that the eigenvalue closest to zero in the pseudospectrum, deep in the topological regime, settles at $-1/2$, rather than $0$ as in the energy spectrum. We can understand this by examining the eigenstates corresponding to those values in the spectrum. Deep in the topological regime, we expect the eigenstates of $H$ close to zero energy to be nearly identical to the eigenstates of $N$ and should have a pseudospectrum eigenvalue corresponding to the contribution from $X\Gamma$ alone. For a state concentrated at the boundary, this contribution should be $-1/2$. In the trivial regime, disorder leads to localized states for all energies. Consequently, the eigenvalues and eigenstates of $N,H$ should be similar to one another, with the eigenstates not necessarily located at the boundary. This also explains why in the trivial phase, the signature of $N$ is identical to that of $H$, which is zero.
\end{widetext}
\end{appendix}

\newpage{\pagestyle{empty}\cleardoublepage}

\end{document}